\documentstyle[12pt]{article}
\begin{document}
\begin{titlepage}
\begin{flushright}
SU-TP-1/95
\vspace{1 cm}
\end{flushright}
\begin{center}
{\LARGE\bf Inverse neutrinoless double-$\beta$ decay \\ in gauge theories
with CP violation
\footnote{This work was supported by Polish Committee for Scientific Researches
under Grant No. 2P30225206/93 and University of Silesia internal Grant.}}  \\
\vspace{1.5 cm}
{\large\bf J.Gluza}$^{\dagger}$ and {\large\bf M.Zra{\l}ek}$^{\ddagger}$ \\
\vspace{ 0.5 cm}
Department of Field Theory and Particle Physics \\
Institute of Physics, University of Silesia \\
Uniwersytecka 4, PL-40-007 Katowice, Poland \\
\vspace{1cm}
\baselineskip 1 mm
{\large\bf Abstract} \\
\end{center}
{\small We investigate the $e^-e^- \rightarrow W_i^-W_j^-$ reactions for the
various
gauge bosons production processes in the frame of the standard model with the
additional right-handed neutrinos and in the Left-Right symmetric model. The
present bounds on the various model parameters are taken into account. The
question of the cross section behaviour for large energy and the CP violation
problems are discussed.}

\vspace{0.5 cm}

PACS number(s):13.15.-f,12.15.Cc,11.30.Er

\end{titlepage}

\baselineskip 6 mm
\section{Introduction}
The nature of the known electron, muon and tau neutrinos and the existence of
the heavier
($M_{\nu} > M_Z$) neutrinos are still open problems in the particle physics.
In the standard model all neutrinos are massless. There is, however, some
more or less strongly established evidence which requires the massive
neutrinos. Problems
with (i) the Sun (electron) neutrinos, (ii) atmospheric (muon) neutrinos and
(iii) the hot dark matter can be resolved if neutrinos are massive particles.
Within the framework of the extended electroweak models neutrinos are usually
massive and Majorana-type. At low energies such models can be
probed by looking for rare processes, such as the neutrinoless double-$\beta$
decay. \\
High energy $e^-e^-$ collision, a possible option in Next Linear Colliders,
may provide a new test for those $\Delta L=2$ interactions via the
$e^-e^- \rightarrow W^-_i W^-_j$ reaction, where $W_i$ may represent
standard model gauge bosons or the additional charged gauge bosons. This
inverse neutrinoless
double-$\beta$ decay was proposed some time ago [1] and since then
has been investigated several times [2]. We have decided to study the process
once more for a few reasons. First, we would like to give the numerical
values of the total cross sections for the $e^-e^- \rightarrow W^-_iW^-_j$
processes, taking into account the up-to-date bounds for the various model
parameters.
Secondly, we would like to study the problem of unitarity
and to find the conditions for correct high-energy behaviour of the cross
sections in various gauge models.
For this purpose it is necessary to consider the all model's ingredients (for
example, in
the L-R symmetric model both left-, and right-handed double-charged gauge
bosons
$\delta_{L,R}^{\pm \pm}$ have to be considered).
The calculated helicity amplitudes for the
process give us the opportunity to have a look at the unitarity cancellations
in a very precise way. Finally, we investigate the problem of
CP-symmetry breaking in our proces and compare its size for various models. \\
In the next Section we define the mass lagrangian and the CP breaking
parameters. In Section 3 we investigate which ingredients of the models
are responsible for correct high-energy cross section behaviour. The numerical
values of total cross sections for model's parameters, which satisfy
existing bounds and the size of CP-symmetry breaking, are given in Section 4.
\section {The mass lagrangian and CP violation}

In the class of models which we consider the mass Lagrangian
for neutrinos and charged leptons is given by
\begin{equation}
L_{mass}=-\frac{1}{2}(\bar{N}_L^c M_{\nu}N_R+\bar{N}_R M_{\nu}^{\ast} N_L^c)-
(\bar{l}_L M_l l_R+\bar{l}_R M_l^{\dagger} l_L),
\end{equation}
where
$$
N_R= \left( \matrix{ i\gamma^2 \nu_L^{\ast} \cr
                     \nu_R } \right) ,\;\;
N_L^c= \left( \matrix{ \nu_L \cr
                     i\gamma^2 \nu_R^{\ast} } \right)
$$
are $n_L+n_R$ dimensional row vectors of neutrino fields and $l_{L(R)}$
are $n_L$ dimensional charged lepton fields. $M_l$ and $M_{\nu}$ are
$n_L \times n_L$ complex and
$(n_L+n_R) \times (n_L+n_R)$ symmetric-complex matrices respectively.
The matrix $M_{\nu}$ is usually divided into four parts
\begin{equation}
M_{\nu} =
{ \overbrace{M_L}^{n_L} \ \overbrace{M_D}^{n_R} \choose M^T_D \ M_R }
{\begin{array}{c}
\} n_L \\ \} n_R
\end{array}}.
\end{equation}

We know, that without changing the physical meaning (all elements of the
lagrangian, except the mass term (Eq.(1)), will not change) some matrices
can be made diagonal.
In the models without right-handed charged current
interaction two matrices, e.g. $M_l$ and $M_R$ can be chosen diagonal,
in the L-R symmetric models only one e.g. $M_l$.
In all kinds of models which we consider the basis of lepton fields can be
chosen in such a way that the charged lepton mass matrix $M_l$ is real and
diagonal.
Then all the lepton-violating CP phases are present in the neutrino mass matrix
$M_{\nu}$ (Eq.(2)). The number of CP-breaking phases depends on
the model. As an example we consider two kinds of models:
\newline
$ \underbar{(1) Standard model with the additional three neutral Right-Handed
Singlets (RHS).} $

In this model $n_L=n_R=3$ and in our basis
\begin{equation}
M_{\nu}= \left( \matrix{ 0 & M_D \cr
                       M_D^T & M_R } \right)
\end{equation}
where $M_R=\mbox{\rm Diag}\left( M_1,M_2,M_3 \right) $
are real numbers and
\begin{equation}
M_D= \left( \matrix{ a_1e^{i\alpha_1} & a_2e^{i\alpha_2} & a_3e^{i\alpha_3} \cr
                  b_1e^{i\beta_1} & b_2e^{i\beta_2} & b_3e^{i\beta_3} \cr
                  c_1e^{i\gamma_1} & c_2e^{i\gamma_2} & c_3e^{i\gamma_3} } .
\right)
\end{equation}
The CP-symmetry is satisfied, if [3]
\begin{equation}
\sin{(\alpha_i-\alpha_j)}=\sin{(\beta_i-\beta_j)}=\sin{(\gamma_i-\gamma_j)}=0
\;\;\;\;\mbox{\rm for}\;\;i,j=1,2,3
\end{equation}
and six phases $\alpha_1-\alpha_2=\chi_1,\; \alpha_1-\alpha_3=\chi_2,\;
\beta_1-\beta_2=\rho_1,\; \beta_1-\beta_3=\rho_2,\;
\gamma_1-\gamma_2=\eta_1$ and $\gamma_1-\gamma_3=\eta_2$ break the symmetry.
\\
\newpage
$ \underbar{(2) The Left-Right symmetric model.} $

We consider two versions of the model. At the beginning we assume that
there is the explicit L-R symmetry with all vacuum expectation values real
(no spontaneous CP-symmetry breaking) [3]. Then the $M_L$ and $M_R$ matrices
are proportional
\begin{equation}
M_L=\mbox{\rm const.} \times M_R
\end{equation}
and $M_D$ is hermitian
\begin{equation}
M_D=M_D^{\dagger}.
\end{equation}
There are only six CP-violating phases [3] and we can choose them in the
following way
\begin{eqnarray}
M_R&=&\left( \matrix{ M_{11}e^{i\alpha_1} & M_{12} & M_{13} \cr
                    M_{12} & M_{22}e^{i\alpha_2} & M_{23} \cr
                    M_{13} & M_{23} & M_{33}e^{i\alpha_3} } \right)
\nonumber \\
\mbox{\rm and}\;\;\;\;\;\;\;\;\;\;\;\;\;\; &&\\
M_D&=&\left( \matrix{ m_{11} & m_{12}e^{i\beta_1} & m_{13}e^{i\beta_2} \cr
                    m_{12}e^{i\beta_2} & m_{22} & m_{23}e^{i\beta_3} \cr
                    m_{13}e^{i\beta_2} & m_{23}e^{i\beta_3} & m_{33} } \right).
\nonumber
\end{eqnarray}

This model is known in the literature as Manifest or Quasi-Manifest
L-R Symmetric one (MLRS or QMLRS). \\
We consider also the full version of the L-R model where
there is no relation between $M_L$ and $M_R$ matrices and $M_D$ is not
hermitian (Non-Manifest L-R Symmetric model - NMLRS). In this case there are
18 CP-violating phases. The phases of the matrices $M_L,\;M_R$ and $M_D$,
which do not satisfy the CP-conserving relations [3],
\begin{eqnarray}
(M_L)_{ij}&=&\mid (M_L)_{ij} \mid e^{i\frac{1}{2}(\delta_i+\delta_j)},
\nonumber \\
(M_R)_{ij}&=&\mid (M_R)_{ij} \mid e^{-i\frac{1}{2}(\delta_i+\delta_j)}, \\
\mbox{\rm and }\;\;\;\;\;\;\;\;\;\;\;\;\;\;\;\;\;\; && \nonumber \\
(M_D)_{ij}&=&\mid (M_D)_{ij} \mid e^{-i\frac{1}{2}(\delta_i-\delta_j)}
\nonumber
\end{eqnarray}
break the CP symmetry. \\
The neutrino mass matrix is diagonalized by the complex orthogonal
transformation
\begin{equation}
U^TM_{\nu}U=diag[\mid m_1 \mid ,\mid m_2 \mid ,...,\mid m_{n_L+n_R} \mid ]
\equiv M_{\nu d}.
\end{equation}
If we denote
$$U= \left( \matrix{ U_L^{\ast} \cr
                    U_R } \right)
$$
then the mixing matrices $K_L$ and $K_R$ in the physical left- and
right-charged
current interactions (see Eq.(A2) in the Appendix) are given by
\begin{equation}
K_{L,R}=U_{L,R}^{\dagger}.
\end{equation}
There are reversed connections between $M_{L(R)},M_D$ and $M_{\nu d}$
\begin{eqnarray}
M_L&=&K_L^{\dagger}M_{\nu d}K_L^{\ast},\;\;\;M_D=K_L^{\dagger}M_{\nu d}K_R ,
\nonumber \\
\mbox{\rm and} \;\;\;\;\;\;\;\;\;\;\;\;\;\; && \\
M_R&=&K_R^TM_{\nu d}K_R \nonumber
\end{eqnarray}
which will be useful in the further considerations.
\section{Behaviour of cross sections for high energy.}

The reduced helicity amplitudes (see the Appendix) have bad
high energy behaviour. As $\lim_{s \rightarrow \infty } \gamma_{1,2} =
\frac{\sqrt{s}}{2M_{1,2}}$ the amplitudes with
$\mid \Delta\lambda \mid=1$ proportional
to $\gamma_1$ and the amplitudes with $\lambda_1=\lambda_2=0$ (proportional
to $\gamma_1\gamma_2$) icrease with increasing energy. Of course, these
divergences
can't appear in the total cross section so there must exist mechanisms
which cause their cancellation.
As the cancellations of these divergences have influence on the size
of the total cross section it is instructive to show how it happens.
There are several reasons why the bad high energy behaviour does not appear
in the helicity amplitudes
for $\sqrt{s} \gg m_a,M_i$,  a=1,...,6,  i=1,2. \\
In the L-R symmetric model for $\Delta\sigma=\pm 1$ in the t- and the
u-channels there
are (see the Appendix)
\begin{equation}
\begin{array}{l}
R_{t(u)}M_{t(u)}\left( \Delta\sigma=\pm 1, \mid \Delta\lambda \mid =1 \right)
\longrightarrow \\
\;\;\;\;\;\;\;\;\;\;\;\;\; \frac{1}{M_i\sqrt{s} \left( 1 \mp \cos{\Theta}
\right)} \sum_{a} \left( K_{L(R)}^T \right)_{ma}m_a^2 \left( K_{R(L)}
\right)_{an}
\end{array}
\end{equation}
and
\begin{equation}
\begin{array}{l}
R_{t(u)}M_{t(u)}\left( \Delta\sigma=\pm 1, \lambda_1=\lambda_2=0 \right)
\longrightarrow  \\
\;\;\;\;\;\;\;\;\;\;\;\;\; \frac{1}{2M_iM_j(1\mp \cos{\Theta})_a} \sum_{a}
\left( K_{L(R)}^T \right)_{ma} m_a^2 \left( K_{R(L)} \right)_{an}
\end{array}
\end{equation}
To obtain these formulae we use the orthogonality property of the $K_R$ and
$K_L$ matrices
\begin{equation}
K_{L(R)}^T K_{R(L)}=0.
\end{equation}
For $\Delta\sigma=0$ the problem is only with one amplitude for
$\lambda_1=\lambda_2=0$. Separately, each amplitude in the u,t and s channels
is divergent but when we add them together, then
\begin{eqnarray}
&&B^{ij}\left( \sigma_1,\sigma_2 \right) [R_tM_t+R_sM_s] +
B^{ij}(\sigma_2,\sigma_1)R_uM_u \longrightarrow  \nonumber \\
&& \nonumber \\
&&\;\;\;\;\; -\frac{\sqrt{2}}{\sqrt{s}M_iM_j}B^{ij}\left( \pm \frac{1}{2},\pm
\frac{1}{2} \right)
\biggl[ \frac{1}{2\left( 1-\cos^2{\Theta} \right) } \sum_{a}
\left( K_{R(L)}^T \right)_{ma} m_a^3 \left( K_{R(L)} \right)_{an} \biggr.
\nonumber \\
&& \nonumber \\
&&\;\;\;\;\; \biggl. + \left( M_{\delta_{R,L}}^2-i\Gamma_{\delta_{R,L}}
M_{\delta_{R,L}} \right)
\sum_{a} \left( K_{R,L} \right)_{ma} m_a \left( K_{R,L} \right)_{an} \biggr] .
\end{eqnarray}
The crucial thing to obtain this high enery behaviour is existence of the three
s,t and u channels. Without the
$\delta_{R,L}^{--}$ bosons exchange in the s channel the unitarity would be
violated. And it is important that both left $\delta_L^{--}$ and right
$\delta_R^{--}$ doubly-charged Higgs bosons are present. They give
contributions to the different helicity amplitudes (see Appendix). \\
We can see that in the models with L-R symmetry the appropriate high energy
behaviour is guaranteed because of the following reasons: \\
(i) the left and right mixing matrices in the charged current are orthogonal
(Eq.(15)), \\
and \\
(ii) there exist two doubly-charged Higgs bosons $\delta_{L,R}^{--}$ with
proper relations between various couplings. \\
We can ask now how it is possible, that in the models without L-R symmetry,
where only one charged gauge boson $W^{\pm}$ exists, the unitarity is also
satisfied. In these class of models, however,  there is no right-handed charged
current so
the mixing angle $\xi$ and the mixing matrix $K_R$ vanish
\begin{equation}
\xi =0,\;\;K_R=0,
\end{equation}
and there are no doubly charged Higgs bosons so there is no contribution to the
amplitude in the s-channel
\begin{equation}
M_s=0.
\end{equation}
Only one helicity amplitude in the t- and u-channels $M_{t,u}(-\frac{1}{2},
-\frac{1}{2},0,0)$ seems to be divergent. But if we look carefully at the high
energy behaviour we have
\begin{eqnarray}
R_{t(u)}M_{t(u)}\left( -\frac{1}{2},-\frac{1}{2},0,0 \right) & \longrightarrow
&
\frac{1}{\sqrt{2}M_iM_j} \biggl[ \sqrt{s} \sum_{a} \left( K_L^T \right)_{ma}
m_a
\left( K_L \right)_{an} \biggr. \nonumber  \\
&&+ \frac{1}{\sqrt{s}} \biggl. \frac{1}{\left( 1-\mp\cos{\Theta} \right) }
\sum_{a}
\left( K_L^T \right)_{ma} m_a^3 \left( K_L \right)_{an} \biggr] \nonumber
\end{eqnarray}
\begin{equation}
\end{equation}
Now from Eq.(12) there is
\begin{equation}
\sum_{a} {\left( K_L^T \right)}_{ma} m_a {\left( K_L \right)}_{an} =
{\left( K_L^T M_{\nu d} K_L \right)}_{mn}
= \left( M_L^{\ast} \right)_{mn}
\end{equation}
In the RHS models without doubly-charged bosons the appropriate Yukawa
mechanism which generates the mass matrix $M_L$ is not present, so
\begin{equation}
M_L=0,
\end{equation}
and the first term in Eq.(19) which is proportional to $\sqrt{s}$ disappears,
what guarantees
the correct
high energy behaviour of the cross section.
In the models where Higgs triplets are present the unitarity is preserved
in similar way as in the L-R symmetry models.

\section{Numerical results}

First we investigate the total cross sections for production of gauge bosons
in $e^-e^-$ reaction and their dependence on various model parameters.
Then we consider the CP symmetry breaking.
\subsection{Total cross section in L-R model.}

To calculate the cross section we need to know the values of the model
parameters (see [4]): \\
\begin{itemize}
\item the mixing angle $\xi$
\begin{equation}
\xi \simeq \left( \frac{2k_1k_2}{k_1^2+k_2^2} \right) \frac{M_{W_1}^2}
{M_{W_2}^2};
\end{equation}
\item the masses of the gauge bosons
\begin{equation}
M^2_{W_1} \simeq \frac{g^2}{2} \left( k_1^2 + k_2^2 \right),\;
M^2_{W_2} = \frac{g^2}{2} v_R^2;
\end{equation}
\item the masses of doubly-charged Higgs particles
\begin{equation}
M^2_{\delta_L} \simeq \frac{1}{2}v_R^2\;,\;\;
M^2_{\delta_R} \simeq 2 v_R^2.
\end{equation}
\end{itemize}
In our numerical analysis we take that $2k_1k_2 \simeq k_1^2+k_2^2$, so
practically only one parameter, the mass of heavy gauge boson
$M_{W_2}$ is free ($M_{W_1}$ and $g=e/\sin{\Theta_W}$ are known from the
standard model).
We do not calculate the decay width for doubly-charged Higgs bosons
but we put them in the form
\begin{equation}
\Gamma_{\delta_{L,R}^{--}}=\Gamma_{W_1}M_{\delta_{L,R}^{--}}/M_{W_1}.
\end{equation}
Besides the above model parameters the cross section depends very
strongly on the neutrino masses and mixings.
First of all, we can see (Eqs.(15),(A6)-(A8)) that if all neutrinos are
massless
($m_a=0$), then
the functions $R_{t,u,s}$ vanish and the cross section is zero. This fact
is very well known. For massless neutrinos there is no way to distinguish
between
the Dirac and the Majorana cases, lepton number conservation is restored and
our
process can not occur\footnote{Strictly speaking it is imposible to distinguish
between the Dirac
and Majorana neutrinos for $m_a \rightarrow 0$ if there are only left handed
or right handed current interactions. If both couplings are present (as in the
L-R
symmetric model) the massless Dirac and Majorana neutrinos are still
indistinguishable
because of the orthogonality of $K_L$ and $K_R$ matrices
$(K_{L,R}^TK_{R,L}=0)$.}.

{}From existing terrestrial experiments we know that [5]
\begin{equation}
m_{{\nu}_e} < 5.1 eV,\;\;
m_{{\nu}_{\mu}} < 0.27 MeV,\;\;
m_{{\nu}_{\tau}} < 31 MeV
\end{equation}
and the heavy neutrinos, if they exist, must have masses $M_N>\frac{M_Z}{2}$
[5] or even $M_N > M_Z$ if additional assumptions about the $\nu NZ$
coupling are made [6].
There are other laboratory experiments (double-$\beta$ decay and neutrino
oscillation in the vacuum) which tell us something about neutrino masses and
mixings. Moreover, there are also solar, astrophysical and cosmological
observations which can also give some information about masses and mixings,
see e.g. [7].
It follows from all existing observations that the three known neutrinos
 $( \nu_e ,\;\nu_{\mu},\;\nu_{\tau})$ should have small, almost degenerate
masses in the range between 0 and several eV. The other three neutrinos
predicted
by the L-R model must have masses above $\sim 100$ GeV.

One of the
possible choices to assure the mentioned pattern of the neutrino masses
is to assume that $M_D$ is almost a rank 1 matrix and $M_R$ is almost
diagonal with large diagonal elements $M_i$ (i=1,2,3) which satisfy the
constraints [8]
\begin{equation}
\frac{1}{M_1}+\frac{1}{M_2}+ \frac{1}{M_3} \simeq 0.
\end{equation}
On the other hand the elements of the $M_R$ matrix are determined by the
right-handed vacuum expectation value $v_R$ and should be bounded by mass
of the heavy gauge boson (Eq.(23))
\begin{equation}
M_i \leq \frac{2M_{W_2}}{g}.
\end{equation}
Let us take the matrices $M_D$ and $M_R$ in the form
\begin{equation}
M_D=\left( \matrix{ 1.0 & 1.0 & 1.0 \cr
                    1.0 & 1.0 & 1.0 \cr
                    1.0 & 1.0 & 1.0-10^{-6} } \right) \;
\mbox{\rm and}\;\;\;
M_R=\left( \matrix{ M_1 & 10^{-6} & 10^{-6} \cr
                    10^{-6} & M_2 & 10^{-6} \cr
                    10^{-6} & 10^{-6} & M_3 } \right).
\end{equation}
Then one neutrino is massless $m_1=0$, the other two have very small masses,
 $m_2\simeq 0,\;m_3 \simeq 0 $ and the masses of the heavy neutrinos are given
by
the diagonal elements of matrix $M_R$
$$ m_4 \simeq \mid M_1 \mid ,\;m_5 \simeq \mid M_2 \mid ,\;
m_6 \simeq \mid M_3 \mid .$$
For this choice of the neutrino mass matrix the total cross section for various
other model parameters is presented on Fig.2.
In the frame of NMRLS model the present experimental bound on $M_{W_2}$ is no
so high and $M_{W_2} \geq 600$ GeV is still a reasonable limit [9]. Then the
mixing angle $\xi$ does not suppress so much the helicity amplitudes.
Curves depicted by capital A present the result for $M_{W_2}=600$ GeV.
In the frame of MLRS or QMLRS models the bound on $M_{W_2}$ is larger
$M_{W_2} \geq 1600$ GeV [10] and the cross sections are lower (the line
depicted by capital B on Fig.2). The reasonably
high luminosity in the next $e^-e^-$ collision will not give the possibility
to observe  the $W_1^-W_1^-$ pair production in this case.
The cross
section depends also on the heavy neutrinos masses. For $m_4 \sim 100$ GeV
(and $m_5\sim m_6 \sim 200$ GeV) the cross section is small (dashed lines on
Fig.2) and increases with the neutrino mass (the solid lines give the cross
section for biggest acceptable values (Eq.(28)). On the figure we can also see
the
influence of the right-handed resonance $\delta_R^{--}$, as for $M_L \sim 0$
the effect of the left-handed resonance $\delta_L^{--}$ is not visible (see
final remarks in the Appendix).

On the next figures we present the cross sections for production of the
light-heavy $W_1^-W_2^-$ gauge bosons (Fig.3)
and two heavy $W_2^-W_2^-$  gauge bosons (Fig.4). The cross sections are higher
but of course the thresholds for these production processes are too high to
hope that the appropriate colliders will be built in the near future. \\
Fig.5 gives the behaviour of the total cross section for two light gauge
boson production $e^-e^- \rightarrow W_1^-W_1^-$ as a function of heavy
neutrino masses. We take the following parametrization of the
neutrino mass matrix in Eq.(29)
\begin{equation}
M_1=-M,\;\;M_2=M_3=2M.
\end{equation}
The behaviour of $\sigma \left( M \right) $ agrees with our previous
discussion:
$\sigma \rightarrow 0$ for $M \rightarrow 0 $. The cross section
increases with the increasing neutrino masses (not shown on the figure) if we
go
with the mass M above the limit given by Eq.(28).

The dependence of the total cross section on the gauge boson mass $M_{W_2}$,
$\sigma \left( M_{W_2} \right) $, for various CM energies and different mass
matrix
parametrizations is depicted on Fig.6. As was discussed before, the cross
section depends strongly on $M_{W_2}$ and this behaviour is not influenced too
much
by the neutrino
matrix parametrization.
\subsection{Total cross section in the models with RHS.}

We consider the model with three additional right-handed neutrinos which are
singlets of the gauge group. At the beginning we assume that the neutrino mass
matrix is pure real (CP is conserved) and is the same as before (Eq.(29)).
There are no doubly-charged Higgs bosons so only the t and u channels
contribute
to the light gauge boson production amplitude $e^-e^- \rightarrow W_1^-W_1^-$
( $\xi =0$ and $K_R=0$ ).
Only one helicity amplitude with $\sigma_1=\sigma_2
=-1/2$ gives contribution to the total cross section. The appropriate $K_L$
mixing matrix elements, $\left( K_L \right)_{ei}$ (i=4,5,6) are decreasing
functions of the heavy neutrino mass and the same can be observed in the total
cross section (Fig.7). The $\sigma \left( e^-e^- \rightarrow W_1^-W_1^-
\right) $ dependence on the CM energy is presented on Fig.8. The cross section
is smaller than in the L-R symmetric model and decreases with energy.
It is also very sensitive to the neutrino mass matrix parametrization.
If we take for example
\begin{equation}
M_D=\left( \matrix{ 1.0 & 1.0 & 0.9 \cr
                      1.0 & 1.0 & 0.9 \cr
                      0.9 & 0.9 & 0.95 } \right),\;
M_R=\left( \matrix{ M_1 & 10 & 20 \cr
                      10 & M_2 & 10 \cr
                      20 & 10 & M_3 } \right)
\end{equation}
the cross section is much larger (dashed line on Fig.8) and its decreasing
with the energy is smaller, visible only for higher energies (not shown on
Fig.8).
For $\sqrt{s}=500$ GeV $\sigma \simeq .25\;10^{-4}$ fb with parametrization
from Eq.(29) and $\sigma \simeq .5\;10^{-3}$ fb with parametrization
given by Eq.(31).
\subsection{CP symmetry breaking}
There are many phases which can break the CP symmetry (see Eqs.(4,8)). We
don't study the cross section as a function of all of them. We choose
only three phases and we assume that the neutrino mass matrix is given by
Eq.(31),
where the following changes are made
\begin{equation}
M_1 \rightarrow e^{i\alpha}M,\;
M_2 \rightarrow 2e^{i\beta}M,\;M_3 \rightarrow 3e^{i\gamma}M.
\end{equation}
If all other matrix elements in Eq.(31) are real then the phases
$\alpha,\;\beta,\;\gamma$ which are equal 0 or $\pi$, correspond to the CP
conserving
case. \\
If all diagonal elements $M_i$ (i=1,2,3) are real and positive $\left(
\alpha =\beta=\gamma=0 \right)$ then the eigenvalues of the neutrino mass
matrix are also positive and the CP symmetry is conserved if the CP parities of
heavy neutrinos are the same and equal
\begin{equation}
{\eta}_{CP}{\left( N_4 \right)}={\eta}_{CP}{\left( N_5 \right)}=
{\eta}_{CP}{\left( N_6 \right)}=+i .
\end{equation}
The same happenes when all masses $M_1,M_2,M_3$
are real negative $\left( \alpha=\beta=\gamma =\pi \right) $. Then CP is
also conserved if neutrinos CP parities are nagative imaginary
\begin{equation}
{\eta}_{CP}{\left( N_4 \right)}={\eta}_{CP}{\left( N_5 \right)}=
{\eta}_{CP}{\left( N_6 \right)}=-i .
\end{equation}
In both cases above the mixing matrix elements $\left( K_{L,R} \right)_{ei}$,
i=4,5,6 are either pure real for $\alpha=\beta=\gamma=0$ or pure imaginary
if $\alpha=\beta=\gamma=\pi$.
We have checked that for the $e^-e^- \rightarrow W_i^-W_j^-$ processes the
helicity amplitudes with $\Delta\sigma=0$, proportional to the heavy neutrino
masses, are dominant. These helicity amplitudes include either square of the
$K_L$ mixing
matrix elements (for $\sigma_1=\sigma_2=-1/2$) or square of the $K_R$ ones
(for $\sigma_1=\sigma_2=+1/2$) - see Eqs. (A6)-(A8).
That is why, summing the helicity amplitude over neutrino masses (Eq. (A5)), we
get
in both cases mentioned constructive contributions from all of them
(solid line on Fig.9).

In the case of mixing CP parities the CP symmetry is also conserved but the
destructive
interference between contributions from various neutrinos causes that the
cross section decreases. In this case some of $\left( K_{L,R} \right)_{ei}$,
i=4,5,6 matrix elements are real and some are imaginary.

To obtain these CP-violating effects several $K_L$ or $K_R$ matrix elements
must interfere in the same helicity amplitude. The structure of the chosen
neutrino mass
matrix causes that $\left( K_L \right)_{en}$ n=4,5,6
matrix elements are of the similar order
\begin{equation}
\mid \left( K_L \right)_{e4} \mid \simeq \frac{1}{m_4} >
\mid \left( K_L \right)_{e5} \mid \simeq \frac{1}{m_5} >
\mid \left( K_L \right)_{e6} \mid \simeq \frac{1}{m_6}
\end{equation}
but only one suitable element of the $K_R$ matrix is large
\begin{equation}
\mid K_R \mid_{e4} \sim 1 >> \mid \left( K_R \right) _{ei} \mid \;\;\;i=5,6.
\end{equation}
This property causes that only if the $K_L$ matrix elements (Eq.(35))
contribute
to the cross section in the visible way, the CP breaking is seen.
It is just the case for two light gauge bosons $W_1^-W_1^-$ production where
$\cos^2{\xi}$ multiplies the $K_L$ matrix contributions in $\sigma_1=\sigma_2=
-1/2$ helicity amplitude. If the contribution from $\sigma_1=\sigma_2=+1/2$
helicity amplitude (where $\sin^2{\xi}$ is multiplied by $K_R^TK_R$) becomes
important, then the CP symmetry breaking effect decreases. That is why the
effect is visible only outside the $\delta_R^{--}$ resonance region where only
one right-handed mixing matrix $K_R$ gives essential contribution (Eq. (36))
and the interference has no importance.
It also means
that the CP symmetry breaking is more visible for the larger $M_{W_2}$, when
$\sin{\xi} \rightarrow 0$. For the
same reason the CP breaking effect will not be seen in the heavy gauge boson
production ($\cos^2{\xi}$ is multiplied by the $K_R$ matrix contributions)
and weakly visible in the light-heavy gauge bosons production
($\sin{\xi}\cos{\xi}$
multiplies $K_L^TK_L$ ).

If the CP symmetry is violated (phases
$\alpha,\;\beta,\;\gamma \neq 0,\pi$), the cross sections lie
between two limiting lines in Fig.9.

The effects of the CP violation in the standard model with the additional
right-handed-neutrinos are depicted in Fig.10. For the same energy with the
same mass matrix
parametrization the effects are almost as large as those in the L-R symmetric
models.

In contrast to the case of the CP violation in the quark sector where the
effect is small the
violation of the CP symmetry in the lepton sector with the Majorana neutrino
can be
very large. The CP violation phases can change the cross section by even more
than one order of the magnitude.
\section{Conclusions}

We have calculated the total cross section for various gauge bosons production
processes in
$e^-e^-$ scattering. Exact calculations have been done in the frames of two
versions
of the Left-Right symmetric model and the standard model with additional
right-handed neutrinos. We have checked the high energy behaviour for the cross
sections,
$\lim_{s \rightarrow \infty } \sigma (s)$. Having calculated the helicity
amplitudes we have been able to show which ingredients of the models are
responsible
for the correct high energy limit $\sigma (s \rightarrow \infty )=0$. In the
Left-Right symmetric model the unitarity is satisfied because there are s
channels
with doubly-charged Higgs bosons $\delta_{L,R}^{--}$ and the left and
right-handed
mixing matrices are orthogonal. In the standard model with additional RHS
but without Higgs triplets the correct high energy is guaranteed because the
fragment of
the neutrino mass matrix identically equals zero, $M_L=0$. These correct
high energy cancellations cause that the cross section for two light boson
production process
$e^-e^- \rightarrow W_1^-W_1^-$ is small.  We have calculated this cross
section
taking into account existing bounds on the parameters of the considered models.
Optimistically, with not very restrictive bounds on the models parameters, we
have found
that for $\sqrt{s}=500 $ GeV in the L-R symmetric model $\sigma \left(
e^-e^- \rightarrow W^-_1W^-_1 \right) \sim 0.01$ fb and for the standard model
with
RHS
$\sigma \left( e^-e^- \rightarrow W^-_1W^-_1 \right) \sim 0.001$ fb. \\
In the L-R symmetric model the cross sections for production of the
non-standard
gauge bosons $e^-e^- \rightarrow W^-_1W^-_2,W_2^-W_2^-$ are larger but then
also the thresholds for these processes are higher. For $\sqrt{s}=1$ TeV
the cross section  $\sigma \left( e^-e^- \rightarrow W^-(80)W^-(600) \right)
 \sim 10$ fb and for $\sqrt{s}=1.5$ TeV \newline $\sigma \left( e^-e^-
\rightarrow
 W^-(600)W^-(600) \right) \sim 5$ pb. \\
 We have also checked how the CP-violating parameters influence the total cross
 section. We have found that  $\sigma_{tot}$ is the biggest if CP is conserved
and if
 all heavy neutrinos CP-parities $\eta_{CP}$ are the same, equal $+i$ or $-i$.
 If heavy neutrinos have mixing CP-parities (for some $\eta_{CP}=+i$ and
 for others $\eta_{CP}=-i$) or if CP is violated then the cross section is
 smaller. The effect of the CP-symmetry breaking can be very large. This
supports
 the statement that the size of the CP violation can be large if in the lepton
 sector the Majorana neutrinos are present.

\section*{Appendix. Helicity amplitudes for the $l_m^-l_n^- \rightarrow W_iW_j$
process. }
\setcounter{equation}{0}
\renewcommand{\theequation}{A\arabic{equation}}
 Helicity amplitudes for two gauge bosons $W_i^-W_j^-$ (i,j=1,2) production
in two charged leptons $l_m^-l_n^-$ scattering processes
($m,n=e,\mu ,\tau $) are described generally by the Feynman diagrams in
the s,t and u channels (Fig.1). The vertices for the t and u channels
are determined by the charged current lagrangian ($\alpha$ and $\beta$ are
spinor indices)
\begin{equation}
L=\sum_{a=1,...6,b=m,n} \bar{N}_{a\alpha}{\left( {\Gamma_{ab}^{(1)\mu}}
\right)}_{\alpha \beta}
l_{b\beta} W_{\mu}^{(1)}+
\bar{N}_{a\alpha} {\left( {\Gamma_{ab}^{(2)\mu}} \right)}_{\alpha \beta}
l_{b\beta}
W_{\mu}^{(2)}
\end{equation}
where
\begin{eqnarray}
&&{\left( {\Gamma_{ab}^{(1)\mu}} \right)}_{\alpha \beta}=\frac{g}{\sqrt{2}}
\left\{
\cos{\xi} {\left( \gamma^{\mu} P_L \right)}_{\alpha \beta}
{\left(K_L \right)}_{ab}-
\sin{\xi} {\left( \gamma^{\mu} P_R \right)}_{\alpha \beta}
{\left(K_R \right)}_{ab} \right\}, \nonumber \\
&&{\left( {\Gamma_{ab}^{(2)\mu}} \right)}_{\alpha \beta}=\frac{g}{\sqrt{2}}
\left\{
\sin{\xi} {\left( \gamma^{\mu} P_L \right)}_{\alpha \beta}
{\left(K_L \right)}_{ab}+
\cos{\xi} {\left( \gamma^{\mu} P_R \right)}_{\alpha \beta}
{\left(K_R \right)}_{ab} \right\}.
\end{eqnarray}
For the LR model the mixing angle $\xi$ and $6 \times 3$ lepton mixing
matrices $K_{L,R}$ are defined in [4], for example. In the RHS model with $n_R$
right
neutrino singlets $\xi=0\;,\;K_R=0$ and the $K_L$ matrix has dimensions
$(3+n_R,3)$.

In the model where the doubly-charged Higgs bosons $\delta^{++}$ exist
(e.g. models with the Higgs triplets) the considered process has the
s channel diagram too, determined by the $\delta^{++}W^-W^-$ and the Yukawa
lagrangian couplings. In the LR model
\begin{eqnarray}
&&L_{HWW}=-\frac{g^2v_L}{\sqrt{2}}\delta_L^{++} \left( \cos^2{\xi} W_1^- W_1^-
+
\sin^2{\xi} W_2^- W_2^- +\sin{2\xi}W_1^-W_2^- \right)  \nonumber \\
&&-\frac{g^2v_R}{\sqrt{2}}\delta_R^{++} \left( \sin^2{\xi} W_1^- W_1^- +
\cos^2{\xi} W_2^- W_2^- -\sin{2\xi}W_1^-W_2^- \right) + h.c. , \nonumber \\
&& \\
&&L_{Yukawa}=\sum_{a,m,n}\frac{1}{\sqrt{2}v_L} \left[ \delta_L^{++}
 l^T_m CP_L {\left( K_L^T \right)}_{ma}
{ m_a } {\left( K_L \right)}_{an} l_n \right] \nonumber \\
&&+\frac{1}{\sqrt{2}v_R} \left[ \delta_R^{++}  l^T_m CP_R
{ \left( K_R^T \right)}_{ma}
{ m_a } {\left( K_R \right)}_{an} l_n \right] +h.c. \nonumber
\end{eqnarray}
Four Feynman diagrams give contributions to the process $l_m^-l_n^-
\rightarrow W_i^-W_j^-$ at the tree level. The differential cross section
is given by
\begin{equation}
\frac{d\sigma^{ij}_{mn}(\sigma_1,\sigma_2;\lambda_1,\lambda_2)}{d\cos{\Theta}}=
\frac{x}{ 16\pi s}{\mid M(\sigma_1,\sigma_2;\lambda_1,\lambda_2)^{ij}
_{mn} \mid }^2
\end{equation}
where $\sigma_1(\sigma_2)$, $\lambda_1(\lambda_2)$ are the helicities of m(n)
fermions and $W_i(W_j)$ gauge bosons, respectively.
The helicity amplitudes can be written in the form ($\Theta,\;\phi$ are polar
angles of $W_i^-$ in the CM frame)

\begin{eqnarray}
&&M\biggl( \sigma_1,\sigma_2;{\lambda}_1,{\lambda}_2 \biggr)^{ij}_{mn} =
-\frac{e^2}{\sqrt{2} \sin^2{\Theta_W}} \times
D^{J_{max}}_{\Delta \sigma, \; \Delta \lambda} \left(0, \Theta,\phi \right)
\times  \nonumber \\
&&\left\{ B^{ij}(\sigma_1,\sigma_2 )M_t(\Delta\sigma;\lambda_1,\lambda_2)
\left( R_t \right)_{mn}
+B^{ij}(\sigma_2,\sigma_1 )M_u(\Delta\sigma;\lambda_1,\lambda_2)
\left( R_u \right)_{mn} \right. \nonumber \\
&&\left. +B^{ij}(\sigma_1,\sigma_2 )M_s(\Delta\sigma;\lambda_1,\lambda_2)
\left( R_s \right)_{mn} \right\},
\end{eqnarray}
where
\begin{eqnarray}
\left( R_t \right)_{mn} &=&  \sum_{a}
{\left( K_{2\sigma_1}^T \right)}_{ma}
\frac{\left( \frac{m_a}{\sqrt{s}}\right)^{1-\mid \Delta\sigma \mid }}
{A+x\cos{\Theta}-\frac{m_a^2}{2s}}
{\left( K_{2\sigma_2} \right)}_{an},  \\
\left( R_u \right)_{mn} &=&  \sum_{a}
{\left( K_{2\sigma_2}^T \right)}_{ma}
\frac{\left( \frac{m_a}{\sqrt{s}}\right)^{1-\mid \Delta\sigma \mid }}
{A-x\cos{\Theta}-\frac{m_a^2}{2s}}
{\left( K_{2\sigma_1} \right)}_{an},  \\
\left( R_s \right)_{mn} &=& \delta_{\sigma_1,1/2} \frac{1}
{1-\frac{M_{\delta_R}^2}{s}+i\frac{{\Gamma}_{\delta_R} M_{\delta_R}}{s}}
\sum_{a} {\left( K_R^T \right) }_{ma} \left( \frac{m_a}{\sqrt{s}} \right)
{\left( K_R \right)}_{an} \nonumber \\
&+&\delta_{\sigma_1,-1/2} \frac{1}
{1-\frac{M_{\delta_L}^2}{s}+i\frac{{\Gamma}_{\delta_L} M_{\delta_L}}{s}}
\sum_{a} {\left( K_L^T \right) }_{ma} \left( \frac{m_a}{\sqrt{s}} \right)
{\left( K_L \right)}_{an} ,
\end{eqnarray}
and
\begin{eqnarray*}
K_\kappa=\left\{
\begin{array}{cl}
K_R & \kappa=+1 \\
& \\
K_L & \kappa=-1
\end{array}
\right.\;,\;
\end{eqnarray*}
\begin{eqnarray*}
&&\Delta \sigma =\sigma_1-\sigma_2\;\;,\;\; \Delta \lambda =
\lambda_1-\lambda_2\;\;,\;\;
J=\mbox{\rm max}(\mid \Delta \sigma \mid, \mid \Delta \lambda  \mid ), \\
&& \\
&&A=\frac{M^2_{W_i}-M^2_{W_j}}{4s^2}-\frac{1+4x^2}{4}, \\
&& \\
&&x=\frac{k}{\sqrt{s}}\;\;,\;\;
k=\frac{1}{2\sqrt{s}}\sqrt{s^2-2s(M_1^2+M_2^2)+(M_1^2-M_2^2)^2}.
\end{eqnarray*}
The factors B are different for various gauge bosons productions
\begin{equation}
B^{ij} \left( \sigma_1,\sigma_2 \right) =\left\{
\begin{array}{cl}
-2\sigma_1 {\left( \sin{\xi} \right) }^{\mid \mid \Delta \sigma
\mid -2\sigma_2 \mid} {\left( \cos{\xi} \right) }^{\mid \mid \Delta \sigma
\mid -2\sigma_1 \mid } & i=1\;,\; j=2, \\
& \\
4\sigma_1 \sigma_2 {\left( \sin{\xi} \right) }^{(1+\sigma_1 +\sigma_2 )}
{\left( \cos{\xi} \right) }^{(1-\sigma_1 -\sigma_1)} & i=1\;,\;j=1, \\
& \\
{\left( \sin{\xi} \right) }^{(1-\sigma_1-\sigma_2)}{ \left( \cos{\xi} \right)
}^
{(1+\sigma_1+ \sigma_2)} & i=2\;,\;j=2.
\end{array}
\right.
\end{equation}

The reduced helicity amplitudes $M_{t(u,s)}\left( \Delta\sigma;\lambda_1,
\lambda_2\right) $ are gathered in the Table 1, where we use
the following notation
\begin{eqnarray*}
&&y_{1(2)}=\frac{E_{1(2)}}{\sqrt{s}}\;\;,\;\; \beta_{1(2)}=\frac{k}{E_{1(2)}}
\;\;,\;\;\gamma_{1(2)}=\frac{E_{1(2)}}{M_{1(2)}}, \\
&&E_{1(2)}=\frac{s+M_{2(1)}^2-M_{1(2)}^2}{2\sqrt{s}},
\end{eqnarray*}
$E_1(E_2)$ and $M_1(M_2)$ are energies and massess of i(j) gauge bosons
respectively.
The high energy behaviour of the helicity amplitudes are given in the Table 2.
It is worth while to notice several interesting properties of the helicity
amplitudes
(Eq. (A5)):
\begin{itemize}
\item if $M_L=0$ then $\delta_L^{--}$ doesn't contribute to the process
(Eqs.(12),(A8)),
\item the $\delta_L^{--} \left( \delta_R^{--} \right)$ in the s channel
contributes
only to the amplitude with the electron helicities $-1/2(+1/2)$,
\item although there are no Majorana neutrinos in the s channel the amplitude
in this channel is also
proportional to the neutrino massess,
\item the helicity amplitudes with $\Delta\sigma=0$ are proportional to the
neutrino
massess $m_a$.
\end{itemize}

$^{\dagger}$e-mail gluza@usctoux1.cto.us.edu.pl

$^{\ddagger}$e-mail zralek@usctoux1.cto.us.edu.pl
\section*{References}
\newcounter{bban}
\begin{list}
{$[{\ \arabic {bban}\ }]$}{\usecounter{bban}\setlength{\rightmargin}{
\leftmargin}}
\item T.Rizzo,Phys.Lett.B116(1982)23.
\item D.London,G.Belanger and J.N.Ng,Phys.Lett.B188(1987)155; \newline
J.Maalampi,A.Pietil$\ddot{a}$ and J.Vuori,Nucl.Phys.B381(1992)544; \newline
Phys.Lett.B297(1992)327; \newline
T.Rizzo,preprint ANL-HEP-CP-93-24(1993); \newline
C.A.Heusch and P.Minkowski,Nucl.Phys.B416(1994)3; \newline
P.Helde,K.Huitu,J.Maalampi and M.Raidal,HU-SEFT R 1994-09,hep-ph/9409320.
\item P.del Aquilla and M.Zra\l ek, "CP violation in the lepton sector with
Majorana
neutrinos",UG-FT-51/95,SU-TP-2/95.
\item N.G.Deshpande,J.F.Gunion,B.Kayser and F.Olness, \newline
Phys.Rev.D44(1991)837; \newline
J.Gluza and M.Zra\l ek,Phys.Rev.D48(1993)5093;
"Higgs bosons contributions to neutrino production in $e^-e^+$ collisions
in a left-right symmetric model", to appeared in Phys.Rev.D.
\item Particle Data Group,Phys.Rev.D50(1994)1173.
\item L3 Collaboration,O.Adriani et al.,Phys.Lett.B295(1992)371.
\item G.Gelmini and E.Roulet,"Neutrino masses", \newline
UCLA/94/TEP/36,November 1994.
\item G.Ingelman and J.Rathsman,Z.Phys.C60(1993)243.
\item F.Abe et.al.,Phys.Rev.Lett.67(1991)2609.
\item G.Beall,M.Bander and A.Soni,Phys.Rev.Lett.48(1982)848.
\end{list}
\section*{Figure Captions}
\newcounter{bean}
\begin{list}
{\bf Fig.\arabic
{bean}}{\usecounter{bean}\setlength{\rightmargin}{\leftmargin}}
\item  The tree level Feynman diagrams which contribute to the $l_m^-l_n^-
\rightarrow W_i^-W_j^-$ process ({m,n}={e,$\mu$,$\tau$}; {i,j}={1,2}).
In the LR model all three channels are present, while in the RHS model there is
no channel s.
\item The total cross section for the process $e^-e^-\rightarrow W^-_1W^-_1$
(LR model) as a function of the CM energy
for various parametrizations of the neutrino mass matrix. Dashed line is for
the
parametrization given by Eq.(29) with $M_1=-100$ GeV.
Solid line is for the biggest available $M_1$ (Eq.(28)). In all cases $M_2$
and $M_3$ are chosen in such a way that Eq.(27) is satisfied.
\item Process $e^-e^-\rightarrow W^-_1W^-_2$ (LR model) as a function of the CM
energy
for various masses of the heavy neutrinos. Denotations as for Fig.2.

\item Process $e^-e^-\rightarrow W^-_2W^-_2$ (LR model) as a function of the CM
energy
for various masses of the heavy neutrinos. Denotations as for Fig.2.
\item Process $e^-e^-\rightarrow W^-_1W^-_1$ (LR model) as a function of the
heavy
neutrino
mass for NLC, $\sqrt{s}=500$ GeV (dashed line) and $\sqrt{s}=1$ TeV (solid
line) with
parametrization of the heavy neutrino masses by Eq.(29), with
$M_1=-M$, $M_2=M_3=2M$.
\item Process $e^-e^-\rightarrow W^-_1W^-_1$ (LR model) as a function of the
heavy gauge boson
mass for different energies, (a) for $\sqrt{s}=500$ GeV and (b) for
$\sqrt{s}=1000$ GeV. Again dashed and solid
lines are for the neutrino mass paramatrization as in Fig.2.
\item Process $e^-e^-\rightarrow W^-_1W^-_1$ (RHS model)
as a function of heavy neutrino
mass for NLC (dashed line) and 1 TeV (solid line) colliders with
parametrization of the heavy neutrino masses as in Fig.5.
\item Process $e^-e^-\rightarrow W^-_1W^-_1$ (RHS model) as a function of the
CM energy
with
parametrization of the heavy neutrino masses given by Eqs.(29),(30) with
M=100 GeV - solid line and
by Eq.(31) with $M_1=100$ GeV, $M_2,M_3=200$ GeV - dashed line.

\item Process $e^-e^-\rightarrow W^-_1W^-_1$ (LR model) for $M_1=200,M_2=400,
M_3=600$ GeV (Eq.(32)) and $M_{W_2}=1600$ GeV when
CP parity is conserved in the lepton sector.
\item Process $e^-e^-\rightarrow W^-_1W^-_1$ (RHS model) for $M_1=200,M_2=400,
M_3=600$ GeV (Eq.(32)) when
CP parity is conserved in the lepton sector.

\end{list}
\newpage

\begin{table}
\begin{center}
\caption{
The reduced helicity amplitudes for all polarizations of
the $m^-n^- \rightarrow W^-_iW^-_j$ process
($ \tilde{\eta}=\lambda_1 \Delta \sigma ,\; \eta=\Delta \sigma \Delta
\lambda ,\; \check{\eta} =2\sigma_1\lambda_1,\;
\hat{\eta}=\left( \sigma_1+\sigma_2 \right) \left( \lambda_1+\lambda_2
\right) $). }
\begin{tabular}{||c|c||c||c||} \hline \hline
$\lambda_1$
& $\lambda_2$
& $\Delta \sigma = \pm 1$
& $\Delta \sigma = 0 $  \\ \hline
\multicolumn{4}{||c||}{$M_t \left( \Delta \sigma;\lambda_1,\lambda_2
\right) $}
\\ \hline
 1 & 1 & $[ \left( x-\tilde{\eta}y_1 \right) + \frac{\tilde{\eta}
-c}{2}]$ & $\frac{1}{\sqrt{2}} \left( 1+\check{\eta}c \right) $   \\
-1 & -1 &   &    \\ \cline{1-4}
 1 & 0  & $\gamma_2[ \left( 1-\eta \beta_2 \right)
\left( x-\eta y_1 \right) +\eta-c]$ &
 $\frac{1}{\sqrt{2}}\gamma_2 \hat{\eta} \left( 1+\hat{\eta}\beta_2 \right) $ \\
 -1 & 0 & & \\ \cline{1-4}
 0 & 1  & $\gamma_1[ \left( 1-\eta \beta_1 \right)
\left( x+\eta y_1 \right) +\beta_1-c]$ &
 $\frac{1}{\sqrt{2}}\gamma_1 \hat{\eta} \left( 1+\hat{\eta}{\beta_1} \right) $
\\
 0 & -1 & & \\ \cline{1-4}
 1 & -1 & $-\frac{1}{\sqrt{2}}$ & 0 \\
 -1 & 1 & & \\  \cline{1-4}
 0 & 0 & \multicolumn{1}{c||}{$\gamma_1 \gamma_2 \ast$} &
\multicolumn{1}{c||}{$-\frac{1}{\sqrt{2}}\gamma_1 \gamma_2 \ast$}\\
&&
$[y_1 \left( \beta_2 -\beta_1 \right) +x \left( 1-\beta_1 \beta_2 \right)
+\beta_1 -c]$ &
$ \left( 1+\beta_1 \beta_2 - c \left( \beta_1 +\beta_2 \right) \right) $ \\
\hline
\multicolumn{4}{||c||}{$M_u \left( \Delta \sigma; \lambda_1,\lambda_2
\right)$}
\\ \hline
 1 & 1 & $-[ \left( x-\tilde{\eta}y_2 \right) +
\frac{\tilde{\eta}+c}{2}]$ & $\frac{1}{\sqrt{2}} \left( 1-\check{\eta}c \right)
$   \\
-1 & -1 &   &    \\ \cline{1-4}
 1 & 0  & $-\gamma_2[ \left( 1+\eta \beta_2 \right)
\left( x-\eta y_2 \right) +\beta_2+c]$ &
 $-\frac{1}{\sqrt{2}}\gamma_2 \hat{\eta} \left( 1+\hat{\eta}\beta_2 \right) $
\\
 -1 & 0 & & \\ \cline{1-4}
 0 & 1  & $-\gamma_1[\left( 1+\eta \beta_1 \right)
\left( x+\eta y_2 \right) -\eta+c]$ &
 $-\frac{1}{\sqrt{2}}\gamma_1 \hat{\eta} \left( 1+\hat{\eta}{\beta_1} \right) $
\\
 0 & -1 & & \\ \cline{1-4}
 1 & -1 & $-\frac{1}{\sqrt{2}}$ & 0 \\
 -1 & 1 & & \\  \cline{1-4}
 0 & 0 & \multicolumn{1}{c||}{$-\gamma_1 \gamma_2 \ast$} &
\multicolumn{1}{c||}{$-\frac{1}{\sqrt{2}}\gamma_1 \gamma_2 \ast$}\\
&&
$[y_2 \left( \beta_1 -\beta_2 \right) +x
\left( 1-\beta_1 \beta_2 \right) +\beta_2 +c]$ &
$ \left( 1+\beta_1 \beta_2 + c \left( \beta_1 +\beta_2 \right) \right)
$ \\ \hline
\multicolumn{4}{||c||}{$M_s \left( \Delta \sigma; \lambda_1,\lambda_2
\right)$}
\\ \hline
 1 & 1 & 0 &
$2\sqrt{2}$   \\
-1 & -1 &   &    \\ \cline{1-4}
 1 & 0  & 0 & 0 \\
 -1 & 0 & & \\ \cline{1-4}
 0 & 1  & 0 & 0 \\
 0 & -1 & & \\ \cline{1-4}
 1 & -1 & 0 & 0 \\
 -1 & 1 & & \\  \cline{1-4}
 0 & 0 & 0 & $-2\sqrt{2} \left( 1+\beta_1 \beta_2 \right)
\gamma_1 \gamma_2 $ \\ \hline \hline
\end{tabular}
\end{center}
\end{table}

\begin{table}
\begin{center}
\caption{ High energy behaviour of the reduced helicity amplitudes for all
polarizations of
the $ m^-n^-\rightarrow W_i^-W_j^- $ process $\left( \tilde{\eta}=\lambda_1
\Delta \sigma ,\; \check{\eta} =2\sigma_1\lambda_1,\;
\eta=\Delta \sigma \Delta \lambda ,\;
\hat{\eta}=\left( \sigma_1+\sigma_2 \right) \left( \lambda_1+\lambda_2
\right) \right)$.}
\begin{tabular}{||c|c||c||c||} \hline \hline
 $\lambda_1$
& $\lambda_2$
& $\Delta \sigma = \pm 1$
& $\Delta \sigma = 0 $  \\ \hline \hline
\multicolumn{4}{||c||}{$M_t \left( \Delta \sigma;\lambda_1,\lambda_2
\right)$}
\\ \hline
 1 & 1 & $\frac{1}{2}(1-c)$ &
$\frac{1}{\sqrt{2}}(1+\check{\eta}c)$   \\
-1 & -1 &   &    \\ \cline{1-4}
 1 & 0  & $\frac{\sqrt{s}}{2M_2}(1-c)$ &
 $\frac{\sqrt{s}}{2\sqrt{2} M_2} \hat{\eta}(1+\hat{\eta}\beta_2)$ \\
 -1 & 0 & & \\ \cline{1-4}
 0 & 1  & $\frac{\sqrt{s}}{2M_1}(1-c)$ &
 $\frac{\sqrt{s}}{2\sqrt{2}M_1} \hat{\eta}(1+\hat{\eta}{\beta_1})$ \\
 0 & -1 & & \\ \cline{1-4}
 1 & -1 & $-\frac{1}{\sqrt{2}}$ & 0 \\
 -1 & 1 & & \\ \cline{1-4}
 0 & 0 & $\frac{s}{4M_1M_2}(1-c)$ &
 $-\frac{s}{2\sqrt{2}M_1M_2}(1-c)$ \\ \hline
\multicolumn{4}{||c||}{$M_u \left( \Delta \sigma;\lambda_1,\lambda_2
\right)$}
\\ \hline
 1 & 1 & $-\frac{1}{2}(1+c)$ &
$\frac{1}{\sqrt{2}}(1-\check{\eta}c)$   \\
-1 & -1 &   &    \\ \cline{1-4}
 1 & 0  & $-\frac{\sqrt{s}}{2M_2}(1+c)$ &
 $-\frac{\sqrt{s}}{2\sqrt{2}M_2} \hat{\eta}(1+\hat{\eta}\beta_2)$ \\
 -1 & 0 & & \\ \cline{1-4}
 0 & 1  & $-\frac{\sqrt{s}}{2M_1}(1+c)$ &
 $-\frac{\sqrt{s}}{2\sqrt{2}M_1} \hat{\eta}(1+\hat{\eta}{\beta_1})$ \\
 0 & -1 & & \\ \cline{1-4}
 1 & -1 & $-\frac{1}{\sqrt{2}}$ & 0 \\
 -1 & 1 & & \\  \cline{1-4}
 0 & 0 & $-\frac{s}{4M_1M_2}(1+c)$ &
$-\frac{s}{2\sqrt{2}M_1M_2}(1+c)$\\ \hline
\multicolumn{4}{||c||}{$M_s \left( \Delta \sigma; \lambda_1,\lambda_2
\right)$}
\\ \hline
 1 & 1 & 0 &
$2\sqrt{2}$   \\
-1 & -1 &   &    \\ \cline{1-4}
 1 & 0  & 0 & 0 \\
 -1 & 0 & & \\ \cline{1-4}
 0 & 1  & 0 & 0 \\
 0 & -1 & & \\ \cline{1-4}
 1 & -1 & 0 & 0 \\
 -1 & 1 & & \\  \cline{1-4}
 0 & 0 & 0 &
$-\frac{\sqrt{2}s}{M_1M_2}$ \\ \hline \hline
\end{tabular}
\end{center}
\end{table}
\end{document}